\documentclass[12pt]{article}

\usepackage{amsmath}
\usepackage{soul}
\usepackage{amsfonts}
\usepackage{amssymb}
\usepackage{physics}
\usepackage{color, soul}
\usepackage[left=3cm,right=3cm,top=3cm,bottom=3cm]{geometry}
\usepackage{graphicx}
\usepackage[caption=false]{subfig}
\usepackage[indent=0pt]{parskip}
\usepackage{tikz}
\usepackage{array}
\usepackage{booktabs}
\usepackage{xcolor}
\usepackage{times}
\usepackage{float}
\UseRawInputEncoding
\usepackage[numbers,sort&compress]{natbib}

\newcommand{\bp}{\mathbf{p}}

\newcommand{\JP}[1]{\textcolor{black}{#1}}

\newenvironment{sciabstract}{%
\begin{quote} \bf}
{\end{quote}}

\newcounter{lastnote}

\title{Emergent dynamics of active elastic microbeams} 

\author
{Q. Martinet$^{1*}$, Y. I. Li$^{1*}$, A. Aubret$^2$, E. Hannezo$^1$, J. Palacci$^1$\\
\\
\normalsize{$^{1}$ISTA, Institute of Science and Technology Austria}\\
\normalsize{$^{2}$LOMA, Bordeaux, CNRS}\\
\\
\normalsize{$^\ast$To whom correspondence should be addressed; E-mail: jeremie.palacci@ist.ac.at}
}
\date{}

\begin{document}
\baselineskip16pt


\maketitle


\begin{sciabstract}

In equilibrium, the physical properties of matter are set by the interactions between the constituents. In contrast, the energy input of the individual components controls the behavior of synthetic or living active matter. Great progress has been made in understanding the emergent phenomena in active fluids, though their inability to resist shear forces hinders their practical use. This motivates the exploration of active solids as shape-shifting materials, yet, we lack  controlled synthetic systems to devise active solids with unconventional properties. 
Here we build active elastic beams from dozens of active colloids and unveil  complex emergent behaviors such as self-oscillations or persistent rotations. Developing tensile tests at the microscale, we show that the active beams are ultra-soft materials, with large (non-equilibrium) fluctuations.  Combining experiments, theory, and stochastic inference, we show that the dynamics of the active beams can be mapped on different phase transitions which are tuned by boundary conditions. More quantitatively, we assess all relevant parameters by independent measurements or first-principles calculations, and find that our theoretical description agrees with the experimental observations. Our results demonstrate that the simple addition of activity to an elastic beam unveils novel physics and can inspire design strategies for active solids and functional microscopic machines. 

\end{sciabstract}

\subsection*{Introduction}

Active particles inject energy at the individual scale, leading to complex behaviors absent in equilibrium, such as climbing walls  \cite{10.1039/c4sm00975d, 10.1038/s41467-024-45872-1}, following topography \cite{Simmchen:1iy} or responding to shear flows \cite{Anonymous:2012cu,ScienceAdvances.Palacci.2015,10.1038/s41467-022-30611-1}. At higher densities, they constitute active fluids that reveal a rich phenomenology of collective effects: active turbulence \cite{Sokolov:2007lr,10.1103/physrevlett.110.228102,Peng.2021, Sanchez:2013gt,  Adkins.2022,10.1103/physrevx.11.031065}, spontaneous flows \cite{Wu:2017ct}, flocks \cite{Nature.Antoine,10.1103/physrevx.14.031008}, active nematics \cite{Redner:2015eg,Shankar.2022,10.1038/s41467-018-05666-8}  or motility-induced phase separation \cite{Theurkauff:2012jo,Palacci:2013eu, Cates:2015ft}. 
Critically, active fluids flow easily and cannot sustain shear forces, limiting their practical use and engineering applications. This challenge can be overcome by harnessing the flows of active fluids to sculpt the internal architectures of soft  materials \cite{Grober.2023,Frechette} or by engineering active elastic solids, {\it i.e.} solids made of active particles \cite{10.1146/annurev-conmatphys-040821-125506,10.1103/physrevlett.130.130002}. Such active solids are predicted to exhibit distinct properties from active gels or liquids, encompassing systems from biology to robotics.
Chiral crystals made of starfish embryos exhibit odd elasticity \cite{Tan.2022,Scheibner.2020} while active elastic solids of macroscopic robots  or bacterial biofilms present spatiotemporal order \cite{10.1038/s41567-022-01704-x, 10.1038/s41567-022-01836-0}. \JP{Remarkably, the coupling between individual orientational degrees of freedoms and translational ones, {\it i.e} self-aligning active matter \cite{10.1103/revmodphys.97.015007}, as constituents of active solids can lead to emerging dynamics, such as spontaneous oscillations even in overdamped situations \cite{10.1103/physrevlett.130.178202, 10.1088/1367-2630/aa9b48, baconnier2024PRE}. Recently, active solids based on centimetre-scale building blocks showed adaptive locomotion, through emergent feedback loop from non-reciprocal interactions \cite{10.1038/s41586-025-08646-3}.  However, the engineering and study of active solids at the microscale  remain widely open}, notably due to the lack of well-controlled synthetic systems required to engineer novel materials.    
 
In this work, we introduce and study a minimal realization of an active solid: an active elastic beam formed from dozens of active catalytic colloids. 
We first investigate the mechanical properties of the active beams, estimating persistence lengths from shape fluctuations and directly measuring elastic response {\it via} micro-tensile tests. We show that activity results in effectively ultra-soft materials, whose mechanical properties can be tuned by activity. Furthermore, active beams display different classes of complex emergent dynamics that can be tuned by boundary conditions: clamped beams display emergent oscillations, while free beams show a length-dependent transition towards persistent rotation. We develop an analytic model of active elastic beams, based on first principles, which reproduce all of the experimental findings. A salient ingredient of the model is a mutual feedback mechanism between the orientation of active particles and the local velocity of the beams, which we characterize experimentally.  Our theoretical analysis recapitulates that the global dynamics of the active beams are driven by boundary conditions and owe their robustness to phase transitions. 


\subsection*{Assembly and mechanical properties of  active beams}

We assemble linear beams made of 10 to 52 active colloids, {\it i.e} photocatalytic microswimmers that propel at speeds up to 15 $\mu m/s$ \cite{Palacci:2013eu, Aubret:2018cca}. The beams range in length between 6 $\mu$m to 33 $\mu m$ and are characterized hereby by the number  $N=3-17$, where $N+1$ is the number of colloids in the centerline of the beam [Fig.1B-C]. 
We leverage to this end an optical assembly method  developed to assemble metamachines from active colloids \cite{aubret_metamachines_2021}. In brief, we use the interplay between optical forces and self-propulsion to position active colloids in optical traps, set by a spatial light modulator (SLM) [Fig.1A]; the effective diffusivity of the active colloids, enhanced by their self-propulsion ($\sim 75\, \mu m^2/s$), enables rapid occupation of the traps and assembly of the active beams. 
Active beams of increasing sizes and lengths are obtained by extending the template of traps on both ends while continuously trapping the backbone of the structure [movie S1, SI]. \JP{It is notable that the particles of the midline of the structure stand vertically, while the axis of polar propulsion of the lateral lines sits in the plane of the substrate, as visible from the black photocatalytic (hematite) component on the bright-field image [Fig. 1A]. This results in the active beams remaining stable owing to the (non-equilibrium) interactions of the active particles composing the material \cite{aubret_metamachines_2021}, while passive particles would diffuse upon release of the pattern of optical traps.} Remarkably, active beams show rich dynamics [movie S2]. Free beams can exhibit a stochastic or persistent rotation, that appears controlled by the length of the beam. When they are clamped at one end, they present slow oscillations with periods above ten seconds. We  aim to rationalize those emergent properties by first investigating their mechanical properties.  
 
Free active beams exhibit shape fluctuations [Fig.1C], which allow the extraction of the persistence length \cite{lamour_easyworm_2014}  as for biological  and synthetic fibers \cite{pampaloni_thermal_2006,abels_single-molecule_2005,diaz_evaluation_2018}. We measure large persistence lengths, $L_p\sim 1.3\, mm$ over the range accessible in the experiment [Fig.1D][Fig.S3]. Next, we perform a micro-tensile test to directly measure the elastic response of the active beams and use optical forces owing to the softness of the materials. We trap both ends of an active beam using calibrated optical traps and gradually modulate the distance between the traps until the active beam breaks or exits the optical traps [SI, Materials and Methods].  We repeat the  experiment with 8  beams of length N=15  to obtain the stress-strain characterization [Fig.1E]. It shows a linear, elastic response with an ultra-low Young modulus $E\sim 40\, Pa$ in a range of strain below $1\%$, before plastic behavior and rupture of the active beams. Remarkably, the sensitivity of the micro-tensile test allows the quantification of changes in mechanical response by reducing the photocatalytic activity of the colloids of the beam [Fig.1E].
 The effect remains however small owing to the range of activity accessible before the cohesion of the active beams vanishes.   
It is noteworthy that comparably low Young modulus are typical in living systems, such as red blood \cite{mishra_deformation_2014} or  lymphoid cells \cite{darling_high-throughput_2015}, and attained with exotic synthetic materials
\cite{jung_synthesis_2013, silva_acoustic_2019} outside of gels.

In equilibrium, persistence length $L_P$ and Young modulus $E$ are linked.  Thermal agitation $k_B T$ activates fluctuations of shapes that are resisted by  elasticity, prescribing $L_P=E I/k_BT$, where $I$ is the second moment of area of the beam.  From the independent measures of $E$ and $L_P$, and approximating the beam to a continuous slab, $I \approx 2\, \mu m^4 $, we estimate an effective temperature $T^*=EI/(k_\mathrm{B} L_P) \sim  4.7 \times 10^3\, K$,  16 fold over room temperature! We now investigate how this non-equilibrium nature drives the rich dynamics of the active beams [movie S2].


\subsection*{Self-oscillations of clamped active beams}

We consider an active beam (length N=11) clamped through a hexagonal array of optical traps at one end, the opposite end free to move, and quantify its dynamics with high temporal resolution ($30$fps) by measuring its deflection angle $\theta$ [Fig.2A], and the polarity $p(t)$ of the active colloids at the free end, which is expected to drive motion \cite{aubret_metamachines_2021}. \JP{Note that the polarity is defined as the projection of the swimming direction in the direction orthogonal to the beam,  the reasons for which will be explained later.} As expected from the high effective temperature, the angle $\theta$ shows large temporal fluctuations over the course of several minutes, yet with long temporal correlations [Fig.2B]. To gain insight into this anomalous dynamics, we inspect the temporal auto and cross-correlations of the beam angle $\theta$ and active colloid polarity $p$, where  $\JP{R_{X Y}(\tau)}= \expval{X(t), Y(t+\tau)}_t $ is the cross-correlation between $X$ and $Y$ and $\langle ,\rangle$ constitutes a time and ensemble average. 

Remarkably, the autocorrelation function  of  $\theta(t)$, $R_{\theta\theta}$ display periodic peaks for individual beams from $N=7$ to $17$ [Fig.2C], despite extensive noise. Those oscillations have periods typically above ten seconds and \JP{do not vary significantly} for beams of different lengths [Fig. 2D] or activity [SI, Fig. S6]. Importantly, the oscillations are much slower than any other mesoscopic time scales of the system, for instance, the rotational time of active colloids $\tau_r\sim 1 \,s$ (which we estimate from the autocorrelation of $p(t)$, $R_{p p}$ [Fig. 2E], and consistent with previous measurements for individual active colloids). We intuit that this discrepancy of timescales indicates that the self-oscillatory behavior is an emergent feature of the system. This is confirmed by writing down a minimal model for elastic beams subject to active noise and finding that it cannot reproduce even qualitatively our experimental observations [SI Sec.~S8]. 

To summarize, {\it (i)} active beams are ultra-soft active solids ($E\sim 40 \, Pa$), subject to high (active) noise, and  {\it (ii)}  displays complex long-time self-oscillatory behavior that cannot simply be captured by the correlated nature of the active noise of the colloids. This implies that we overlooked a salient physical ingredient to the emergent behavior. Examining the cross-correlation between  $\theta$ and $p$ reveals a complex pattern with both positive and negative cross-correlations at different timescales [Fig. 2F], and hints at a feedback mechanism \JP{we aim to unveil. To this end, we investigate the dynamics of free (untethered) beams, which do not show significant elastic deformations [movie S2]. We intuit that it will enable the direct relationship between the dynamics of the beam and its drive, which we discuss below.
}

\subsection*{Spontaneous rotation of untethered beams from positive hydrodynamical feedback} 

In equilibrium, passive beams are agitated by thermal noise:  the center of mass and orientation exhibit Brownian motion. This contrasts with the dynamics exhibited by the free active beams of all lengths [Movie S2]. Short beams (N=2-5) rotate fast and persistently [Fig.3A], while long beams rotate slower, with stochastic changes of direction [Fig.3B].

We hypothesize that this dynamics is driven by the orientation of the active colloids at the tip, which we test by quantifying the rotational speed of the end of a beam $R\partial_t \theta$ and the \JP{polarity} $p_1, p_2$ of the active colloids at the two ends, denoted as $1$ and $2$. The measurements reveal a strong linear correlation for active beams of all lengths ($N=3-11$) [Fig.3C-inset],  confirming that the rotation is primarily driven by the active colloids at the tip [Fig.S8]. \JP{This is further confirmed by the absence of correlation between the orientation of the swimmers of the side and the dynamics [SI]}. \JP{Because the two ends of the free microbeams are equivalent, we simply denote  as $p$, the polarity $p_{1,2}$.} 
We then quantify the probability distribution function (PDF) of $p$ for beams of different lengths. For short beams that rotate fast, the polarity $p$ is well defined around respectively $-\pi$/4 and $\pi$/4 for clockwise and counterclockwise rotation. With increasing length of the active beams, the rotation rate reduces: the average of $p$ vanishes while its PDF broadens [Fig.3D], as previously observed with colloidal metamachines \cite{aubret_metamachines_2021}. \JP{This interplay between the rotation rate of the microbeam and the polarity $p$ of the driving hints at a self-aligning mechanism, reminiscent of self-aligning polar active matter \cite{10.1103/physreve.84.040301,10.1103/revmodphys.97.015007, 10.1103/physrevlett.130.178202}, which we show of hydrodynamic origin.} We further investigate the implications of this coupling for the dynamics of clamped active beams. We start with a microscopic model of individual microswimmers and employ a combination of fluid dynamics and statistical physics to arrive at predictions for the dynamics of elastic active beams, which we compare to the experiments.

\subsection*{Minimal models of boundary-controlled phase transitions in active beams}

We describe the persistent dynamics of the individual swimmers as an Active Ornstein-Uhlenbeck process (AOUP) \cite{FodorPRL}, whose advantage lies in its analytical tractability. It has been shown that the average dynamics of AOUPs are similar to Active Brownian particles \cite{CapriniSciRep} and undergo Motility-Induced Phase Separation \cite{MartinPRE, LiEPL}. While AOUPs are usually written in terms of self-propulsion velocity, this cannot be conveniently measured in our experiments, where active colloids are tethered to each other. Instead, we express it in terms of polarity \JP{$\bp $}: 
\begin{equation}
    \partial_t  \JP{\bp }= - \gamma \JP{ \bp} + \sqrt{\gamma} \Lambda
    \label{eq:aoup}
\end{equation}
where $\Lambda$ is a spatio-temporal white noise with correlation $\expval{\Lambda_i (t) \Lambda_j (t')} = \delta_{ij} \delta(t-t')$. For single particles without external forces, the polarity $\bp$ undergoes persistent dynamics with a time scale: $1/\gamma$: $\expval{p_i (0) p_j (t)} = \frac{1}{2}   \delta_{ij} e^{-\gamma | t |  }  $\cite{FodorPRL}. Note that on average, \JP{$\bp $} has unit magnitude in 2 dimensions: $\expval{  | \bp (0) |^2} = 1$, though Eq.~\ref{eq:aoup} does not preserve the total magnitude of \JP{$\bp$} at all times. 

Following the linear correlation  between rotation rate and  polarity of the active colloid at each end [Fig.3C-inset], we model the active beam as a passive rigid beam of length $2R$
driven  at each end \JP{$i=1,2$} [Fig.3E]. Force balance \JP{in the tangential direction} requires 
\begin{align}
    \chi_s R^3 \partial_t \theta &= \mu R (p_1 + p_2) + R \JP{\sigma_\mathrm{eff}} \Lambda_\theta  \label{eq:sr_1}
\end{align}
where \JP{$p_1, p_2$ are components of the polarities of the two tip-swimmers in the direction perpendicular to the beam}, $\chi_s R^3 $ represents the frictional moment of inertia, $\mu$ is the force exerted by one swimmer on the beam, $\Lambda_{\theta}$ is a spatio-temporal white noise with unit variance \JP{and $\sigma_\mathrm{eff}$ parameterizes the magnitude of the noise}. Here we take $p_{1, 2}$ to be the projection of swimmer orientation perpendicular to the long axis of the beam, which each follows Eq. \eqref{eq:aoup}. We remark that this leads to a direct proportionality between $\partial_t \theta $ and the mean orientation:  $N^2 \partial_t \theta = s( p_1+p_2)/2 + \text{noise}$, where $s = \frac{ 8 \mu }{ \chi_{s}  l_0^2}$ and $l_0\sim 2a$ is the diameter of the active colloid  \JP{($ R = N l_0 /2 $)} . This prediction is tested against experimental data, revealing a collapse of measurements for spinning beams of all lengths, \JP{where we consider, by symmetry, the average polarity $p$ of the swimmers at the tips} [Fig.3C]. It confirms that the beams are driven by the active colloids at their tips and provides an experimental determination for the parameter $s$. \JP{We directly relate the noise magnitude $\sigma_\mathrm{eff}$ to the effective temperature $T^*$ measured: $\sigma_\mathrm{eff} = \sqrt{2 k_\mathrm{B} T^* \chi_s }$, as further discussed in sec.\ S7, w. }

This simple picture, however, falls short of capturing the long-term dynamics of the system, namely the transition between persistent and stochastic spinning observed for untethered beams as well as the oscillatory dynamics observed for the clamped ones. Indeed, due to the absence of any feedback on the polarity at the tips, $p_i$ ($i=1,2$), it predicts that the beams simply fluctuate with persistence times of order $1/\gamma\sim 1s $, \JP{or  rotational diffusion of the particle at the tip}. We proceed next to investigate the source of feedback that gives rise to the observed long-time behaviors in both free and clamped beams.

In the experiment,  active colloids are sedimented near the bottom substrate of a glass capillary and present a shape asymmetry owing to the hematite cube protruding out of the main spherical body. \JP{As the microbeam moves, the tip experiences a shear flow. Because the particle is asymmetric, the large (spherical) part moves slower than the hematite part in the shear flow, which leads to  the reorientation of the particles at the tip by the shear \cite{LaugaBook}. Such orientation in shear flow, dubbed "rheotactic" behavior, has been previously reported for a wide range of self-propelled agents}, including \textit{E. coli} \cite{HillPRL}, \textit{B. subtilis} \cite{CisnerosAL}, \cite{KantslerElife} and active colloids \cite{ScienceAdvances.Palacci.2015} (we note that this is distinct from the chemically driven flow-polarity coupling in \cite{campbell2019NatComm}). We hereby demonstrate that including this generic feedback of hydrodynamic origin allows us to describe the experimental observations quantitatively. Using Stokes' equation, active colloids kinetically reorient with rotation rate $\omega$ \cite{LaugaBook} [Fig.3E], 
\begin{equation}
\omega = - \nu \dot{\gamma} \sin (\phi)
\label{eq:upstream_reorient}
\end{equation}
where  $\nu$ is a dimensionless \JP{phenomenological} parameter that characterizes the asymmetry of the shape, $\dot{\gamma}$ is the shear rate due to the no-slip at the wall and $\phi$ denotes the angle between the swim direction and the flow. \JP{In practice, the exact value for $\nu$ is set both by the asymmetry of the shape and detailed flow field near the tip. This constitutes a complex hydrodynamic situation arising from the flows generated by dozens of active colloids \cite{10.1021/acsnano.4c18078}, near a boundary, and beyond the scope of the present work. A quantitative analysis of the alignment of  active particles from hydrodynamic shear is presented in [Fig. S7], with  additional experiments at controlled shear.} For simplicity, we approximate the local shear rate $\dot{\gamma}\sim R \partial_t \theta/\xi$  where $\xi$ is the distance between the swimmer and the bottom substrate and $\sin(\phi)  \approx O(1)$, neglecting the dynamics of the component parallel to the beam. Adding this to our model, we obtain:
\begin{equation}
\begin{split}
    \chi_s R^3 \partial_t \theta &= \mu R (p_1 + p_2) + R \JP{\sigma_\mathrm{eff}} \Lambda_\theta  \\
    \partial_t p_i  &= f(  R \partial_t \theta , p_i ) - \gamma p_i + \sqrt{\gamma}\Lambda_i   
    \end{split}
\end{equation} 

\JP{where  $\alpha = \nu/\xi$ for conciseness and $f(  R \partial_t \theta , p_i )$ is the feedback mechanism between velocity and polarity. The hydrodynamic origin of the self-alignment [Eq.3] gives $f(R \partial_t \theta, p) = \alpha R \partial_t \theta$ to lowest order. Polarity $\bp$  is a unit vector, thus requiring the saturation of $p$. We therefore write $f(R \partial_t \theta, p) = \alpha R \partial_t \theta - \beta p^3$, where the $\beta-$term phenomenologically ensures the saturation, while obeying the $(p \rightarrow - p)$ symmetry of the experiment.} 
We unveil the profound consequences of this hydrodynamic coupling of upstream re-orientation on the dynamics of the active beams. 

\subsection*{Supercritical rotation of untethered (or free) active beams}

Taking a mean-field approach, we consider  $p = (p_1+p_2)/2$ and obtain: 
\begin{equation}
    \partial_t p =  \left ( \frac{\bar{\alpha}  }{N }  - \gamma \right )p - \beta p^3  +   \sqrt{ (\gamma/2) }  \Lambda 
    \label{eq:sr_2}
\end{equation}
where $\bar{\alpha} = \alpha s l_0 /  2 $  and the noise variance is $\gamma/2$ for independent Gaussian noises. As  $N$ increases, the linear term goes from positive to negative, marking a supercritical pitchfork bifurcation from two fixed points to one, resulting in the arrest of the persistent rotation for large beams, for which the positive feedback is hindered. The bifurcation is predicted to occur at a critical length N = $\bar{\alpha}/ \gamma $, where the linear term changes sign. For longer beams ($N > \bar{\alpha}/ \gamma)$, we expect to observe stochastic runs whereas fast and persistent rotations are predicted for shorter beams [Fig.3F]. 

We now proceed to compare quantitatively our theory with the experiments via stochastic inference. Since eq.~\eqref{eq:sr_2} is a single-variable stochastic differential equation, we solve for the steady state probability directly \cite{gardiner}: 
\begin{equation}
    - \log P(p) = \frac{4}{\gamma} \left [   \frac{1}{2}  \left (\frac{\bar{\alpha} }{N }  - \gamma \right ) p^2 + \frac{\beta}{4} p^4  \right ] + \text{constant}
    \label{eq:sr_prob}
\end{equation}
The expected steady-state probability is directly compared with the empirical distribution of the orientation $p$ from free beams of 5 different lengths to find the best-fit parameters via stochastic inference [Fig.3G]. The theory shows excellent agreement with the experiment, with the best best-fit parameters recapitulated in [Table 1],  agreeing with values estimated independently with the experiment  [SI \JP{sec.~S7}].  The bifurcation is predicted to occur at a critical length N = 104.3/4 $\sim 26$ in line with the experiments [Fig.3F]. To sum up, our model agrees quantitatively with the experiments.

\subsection*{Hopf bifurcation in the dynamics of clamped beams }

Having successfully described the dynamics of untethered beams, we now turn to the predictions of the model for clamped beams, incorporating the effect of elasticity into the analysis. Using the Euler-Bernoulli equation of beams, and the previous assumptions of a beam driven by the active colloids at the tip [Fig. 4A], we obtain 

\begin{equation}
\begin{split}
   \chi_b  \partial_t \theta(x, t) &= -  E I  \partial_x^4 \theta + \frac{\mu}{L} \delta ( x- L) p  + L^{-1} \JP{ \sigma_\mathrm{eff} } \Lambda_\theta \\
    \partial_t p &= f(L \partial_t \theta |_{x=L}, p) - \gamma p + \sqrt{\gamma} \Lambda_p  \\
    \text{with boundary conditions: } & \theta|_{x=0}= 0, \; \partial_x \theta|_{x=0} = 0, \; \partial_x^2 \theta|_{x=L} = 0, \; \partial_x^3 \theta |_{x=L}=0
\end{split}
\end{equation}
where $\Lambda_{\theta, p}$ are unit white noises, where $E$ is Young's Modulus, $I$ is the second moment of area about the bending axis, and \JP{$\sigma_\mathrm{eff}$ is the effective noise amplitude. As explained in sec.\ S7, we approximate $\sigma_\mathrm{eff} = \sqrt{ 2 k_\mathrm{B} T^* \chi_b }$, where $T^* = 4.7 \times 10^3 K $ is the effective temperature estimated earlier from the independent measurements of the persistence length and Young modulus of the beams.} Additionally, $\chi_b \sim\frac{ 4 \pi \eta}{\sqrt{2 \xi / l_0 }}$ represents the Stokes friction per unit length, where $\eta$ is viscosity and $l_0\sim 2a$ is the diameter of a bead   \cite{Jeffrey}. For convenience, we decompose $\theta(x, t)$ onto the set of eigenfunctions $g_n (x) $ of the bending operator $\partial_x^4$ with clamped boundary conditions  \cite{BauchauBook} [SI Sec.~S8]. We remark that the beams are sufficiently stiff and effectively only exhibit the zeroth mode in the experiment, as further confirmed by comparison of the estimation of eigenvalues from the model: $\lambda_1/ \lambda_0 \approx  81 $. This leads us to only consider the zeroth mode and substituting in the same form of $f$ as before, we have 
\begin{equation}
    \begin{split}
    \partial_t \theta  &= - \epsilon \theta + \bar{\mu} p + \bar{\Lambda}_\theta \\ 
    \partial_t p &= -  (\gamma -   c \bar{\mu}  ) p  - \beta p^3 - c \epsilon  \theta +  \bar{\Lambda}_p 
    \end{split}
    \label{eq:br}
\end{equation}
where $\epsilon =  \frac{E I }{\chi_b L^4} (0.6 \pi ) ^4$ and $\bar{\mu} = \frac{\mu g_0 ( 1) }{\chi_b L^2}$, and $c = \alpha L g_0(1)$ . We refer to [SI Sec.~S9] for further details on the noise correlations. Equations  Eq.~\ref{eq:br} are known as the FitzHugh-Nagumo Model, commonly used to model neuron activities \cite{nagumo1962, fitzhugh1969} and describing a system that can undergo a Hopf bifurcation and admit limit-cycle oscillatory solutions, as shown in [Fig. 4B]. \JP{An equation equivalent to Eq. 8 had been previously derived for active solids of elastically coupled particles that all self-align, and shown to present a Hopf bifurcation as well \cite{baconnier2024PRE}. The present situation differs in that the self-aligning torque is limited to the particles at the tip of the microbeam only.}

Following this preliminary analysis, we turn to a quantitative comparison between the prediction of the clamped beams model, using the best-fit parameters determined previously with free beams as well as independent direct experimental measurements. The previous estimations of the free beam from stochastic inference  [Table 1] provided numerical values for three parameters, roughly corresponding to $\bar{\mu},  c/\gamma, \beta/\gamma$ [Table 2], leaving the 3 parameters $\epsilon, \gamma$ and $\sigma_\theta$ to be estimated from independent measurements.  Using the estimates  $\epsilon = 0.35 s^{-1},  \gamma = 1 s^{-1}$ from mechanical and kinematic experiments, and $\sigma_\theta = 0.03 s^{-1/2}$ [SI Sec. S10], we realize that those parameters place the system in the limit-cycle regime [Fig.4B], explaining the self-oscillations observed experimentally.

In order to compare the aforementioned model with experimental data, we measured the observable $(p,\theta)$ for 14 independent active beams of length $N=11$ [Fig. 4C]. The experimental distributions for $(p, \theta)$ are near Gaussian and remarkably well captured by simulations with the preliminary estimates in [Table 2], [Fig. S12, S13]. Though the preliminary estimates for the parameters set the system in a region of deterministic limit cycles in the phase diagram [Fig. 4B], a limit cycle is difficult to identify experimentally in  $(p, \theta)$ space [Fig. 4C]. In fact, the deterministic limit cycle is obscured by the noise in both $\theta$ and $p$, leading to blurred distribution as confirmed by additional simulations of the model with noise, where the limit cycle is similarly difficult to identify [Fig.4C]. In addition to the observed oscillatory behavior of $R_{\theta\theta}$, subtle differences support the existence of the limit cycles: the peak of the $p$ distribution is slightly flattened compared with the classic bell-shape of Gaussian distributions of the same width, a feature shared by both data and simulations [Fig. 4D]. Finally, we also performed stochastic inference on experimental trajectories of clamped beams, keeping all model parameters \textit{a priori} unknown. Importantly, this gives parameters that were globally consistent with our estimations from first principles as well as stochastic inference from free beams [Table 2], and localizes the system in the same parameter region of limit cycles.

Another signature of Fitzhugh-Nagumo oscillators is the presence of an activator-inhibitor coupling between two variables. In our model, the polarity $p$ behaves as an activator, both promoting rotation (changes in $\theta$) by activity and on itself by upstream re-orientation, while $\theta$, plays the role of inhibitor, restored towards zero by elasticity. We test this in experiments with clamped beams by computing the temporal  cross-correlation function between the two variables, defined as $\expval{ \theta(s) p ( s + t) }_s / \sqrt{ \expval{\theta(s)^2}_s \expval{ p(s)^2 }_s}$. Importantly, the cross-correlation function revealed both a positive peak at $t \sim 0.5s $ (the ``activation'') and a negative peak at $t \sim -0.5s $ (the ``inhibition'') [Fig.4E].   This cross-correlation is dramatically altered in our model when we switch off cross-coupling by either setting $\bar{\mu}$ or $c$ to zero [Fig.S15], implying that {\it both} upstream reorientation ($c$) and self-propulsion of active colloids ($\bar{\mu}$) are critical in the observed behavior of bending beams [SI Sec.~S9.A].  Furthermore, the model predicts autocorrelation functions that agree quantitatively with the experiment [Fig.4F, G]. 
Finally,  we comment on the period of oscillations, which did not change significantly with the lengths explored in the experiment. 
Numerical simulations of the model show a comparable effect with the period of the deterministic oscillations in the experimental range [Fig. 4H]. This provides yet another validation of our model and physical insight into its origin: the length dependence of the parameters is inherently built into the definition of our parameters (see text following \eqref{eq:br}). In effect, parameters do not depend on absolute lengths as for conventional elasticity, providing robust oscillations for beams of various lengths. \\

\subsection*{Discussion}

To conclude, we assembled dozens of active colloids into active beams and investigated their physical properties. The active beams are ultra-soft elastic solids and present emergent dynamics that are controlled by boundary conditions: free beams present a supercritical transition in rotation while tethered beams self-oscillate at a macroscopic time scale of tens of seconds, highlighting in a different context the salient role of boundaries in systems away from equilibrium \cite{NaturePhysics.Solon.2015,NatureCommunications.Bricard.2015,Wu:2017ct}. We compared our experimental measurements with a model of an elastic beam driven by the microswimmers at the tip and presented a quantitative agreement with the experiments, with parameters estimated via a combination of stochastic inference and first principle calculations from direct \textit{independent} measurements. Activity here plays a fundamentally different role compared to previous work on active filaments or biological flagella \cite{10.1063/5.0011466,10.1103/physrevlett.132.248402, 10.1016/j.jmps.2022.104918, 10.1038/s41567-022-01688-8, 10.1063/1.3318497,10.1103/physrevlett.113.048101}.  Firstly, as activity rather than equilibrium interactions are responsible for the cohesion of the beams themselves, we can achieve ultra-soft elasticity. Secondly, the oscillations are an emerging property, rather than a cyclic drive as in most models of flagella. \JP{The system is noticeably different from systems of polar particles linked to each other elastically, where the drive is distributed over the all structure, as beautifully exemplified in \cite{10.1088/1367-2630/aa9b48, 10.1103/physrevlett.130.178202, ferrante2013PRL}.  It therefore remains} to be investigated, whether our synthetic system can lead to collective synchronization as observed for example with metachronal waves \cite{PhysicalReviewLetters.Brumley.2012, 10.1038/s41467-020-16458-4, 10.1073/pnas.1218869110}. Our results will inspire activity-driven  machines that transit between qualitatively different dynamics, controlled by boundary conditions. They demonstrate rich dynamics from simple building blocks and can be leveraged with recent developments in  robotic swarms \cite{10.1126/scirobotics.adn6035} and colloidal robots \cite{10.1038/s41563-023-01589-y}.  Our results show that the interplay of activity and elasticity in solids can emulate bio-inspired functions.

\newpage

\begin{figure}
\centering
\includegraphics[width=\linewidth]{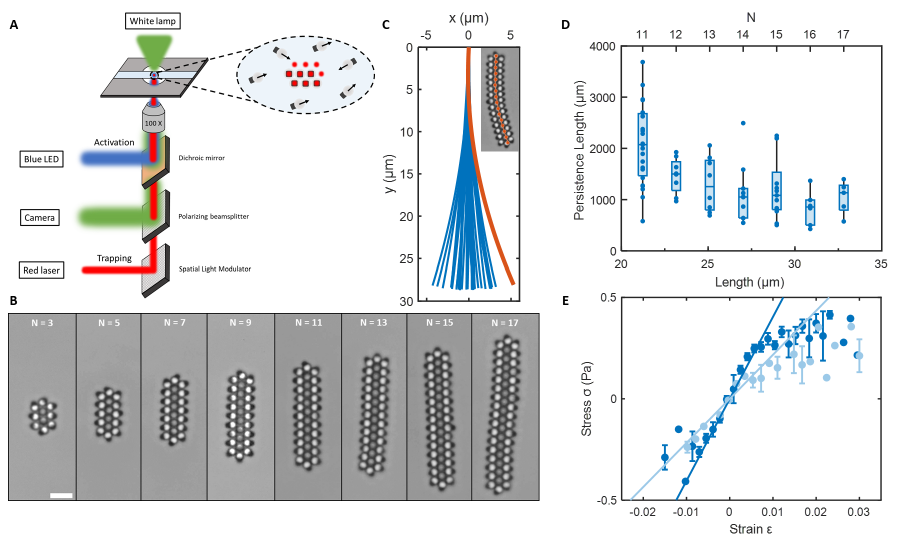}
\caption{{\bf Assembly and mechanical properties of the active beams.} 
(A) Scheme of the optical setup used for the templated assembly of active beams from active colloids. Optical traps (red laser) are controlled by a spatial light modulator, while active colloids are activated by uniform blue light. 
(B) Bright-field images of active beams of increasing lengths, from 6 $\mu$m to 33 $\mu$m. The number (N+1) corresponds to the number of colloids in the central line of the beam. Scale bars are 5 $\mu$m. (C) Time lapse of  the shape fluctuations of an active beam with length 33 $\mu$m. The orange line corresponds to the position tracked in the bright field image in the inset. (D) Measurements of the persistence length $L_p$ for active beams of increasing lengths obtained from the shape fluctuations [see Main text, SI]. 
(E) Stress-strain measurements  obtained by pulling on both extremities of an active beam with optical tweezers [See Main text, SI]. Stress is determined with the optical forces for each trap measuring the trap stiffness and the offset positions of the microswimmers with the optical trap position. Strain is estimated with the relative difference in length of the active beam. Measurements averaged over 8 different beams at maximal activity (dark blue) and $60\%$ activity (light blue). The stress-strain plot shows a linear, elastic regime with Young modulus $E\sim 40Pa$ at maximal activity below $1\%$ strain, before plastic deformation or rupture occurs. Reducing the activity lowers the Young modulus of the active beam.}
\label{fig:figure1}
\end{figure}

\newpage
\begin{figure*}
\centering
\includegraphics[width=\linewidth]{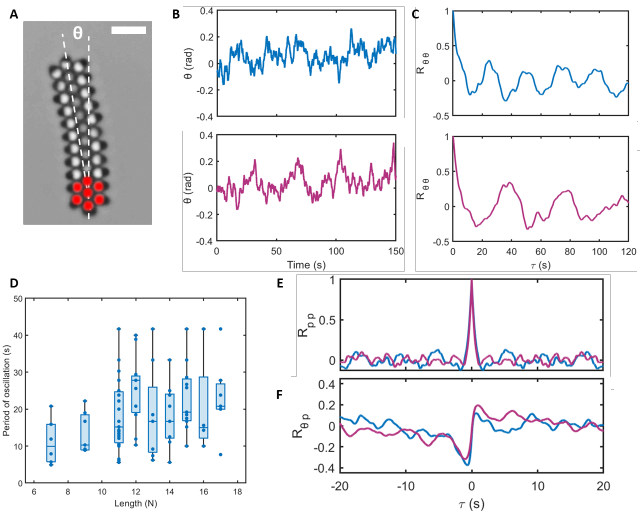}
\caption{{\bf Oscillatory dynamics of clamped  beams.} A) Snapshot of an active beam, with length N = 11, clamped at one end with a hexagonal array of optical traps (red dots). The deformation of the beam is measured with the angle $\theta$. The scale bar is 5 $\mu$m. (B) Evolution of the angle $\theta$ in time for two different beams (N = 11), clamped on one end. The time series are represented in separate panels and different (blue and \JP{magenta}) colors for clarity. (C) Corresponding autocorrelation $R_{\theta\theta} $ of the angle $\theta$, showing periodic peaks and slow oscillations (above ten seconds) and represented in separate panels and colors for clarity.  (D) Periods of oscillations extracted from the Fourier transform of the  autocorrelation $R_{\theta\theta}$ of the angle $\theta$ for increasing lengths of active beams. Periods are typically $\sim 15s$ and do not depend strongly on the length of the beams in the considered range. (E) Autocorrelation of the polarization $R_{pp}$ of the active beams. It shows a rapid decay $\sim 1s$ in agreement  with thermal rotational diffusion. (F) Cross correlations of the bending angle $\theta$ with the polarization $R_{\theta p}$ of the active beams. It shows a usually complex structure with  an autocorrelation that is negative before 0 and positive after, hinting at an activator/inhibitor dynamics [see Main Text].}
\label{fig:figure2}
\end{figure*}

\newpage
\begin{figure*}
\centering
\includegraphics[]{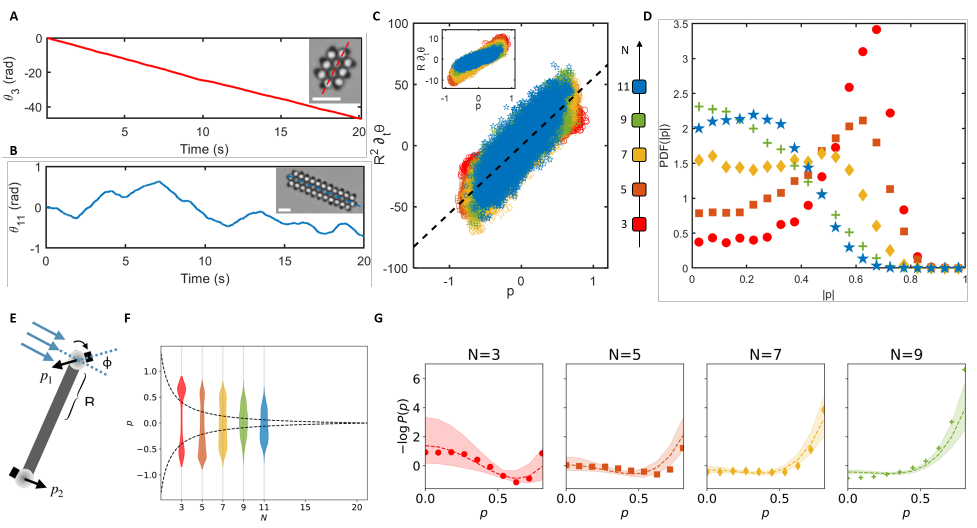}
\caption{{\bf Dynamics of free beams}.  (A) Time evolution of the angle $\theta$ for an active beam of length $N=3$. This short active beam presents a persistent rotation with an angular speed of 2.2 rad/s. (B) Time evolution of the angle $\theta$ of an active beam with length N = 11. This long beam  exhibits stochastic reversal of the direction of rotation. (C-inset) Measurements of the angular velocities $R \partial_{t}\theta$ of active beams of various lengths as a function of their polarization p. Each color corresponds to a different length of the beams (see colorbar). The correlation highlights that the beams are driven by the active colloids on the tip.  (C) Collapse of the experimental data of the inset for the different lengths of beams by plotting $R^2 \partial_{t}\theta$ as a function of the polarization p, as predicted by our model (dashed line) [see Main text, SI]. 
(D) Histogram of the average \JP{absolute} orientation $\mid p\mid $ of the active colloids at the tips for \JP{5} different lengths $N$ of active beams (see colorbar). Distribution broadens and shifts to zero with the length of the beam resulting in a non-persistent rotation.
(E) Schematics of the beam driven by active colloids at two tips. Individual active colloids align their propulsion direction and polarity as a result of the shear flow (blue arrows).  (F) Pitchfork bifurcation of the $p$-dynamics as N decreases (with Maximum Likelihood (ML) estimates), marked by the \JP{dashed} black lines, indicative of the transition from persistent rotation of free beams to stochastic rotation above a critical value $N=26$. The \JP{color coded} violin plots represent the PDF from experimental data from active beams represented on panel D. (G) The effective potential from experimental data (solid circles) overlayed with numerical predictions of the model from Maximum Likelihood estimates using only two parameters (only 4 different lengths N are represented for clarity). The mean of the ML estimates is marked with dashed lines and the errors from bootstrapping statistics are marked with shaded regions of the same color.}
\label{fig:figure3}
\end{figure*}

\newpage
\begin{figure*}
\centering
\includegraphics[width=\textwidth]{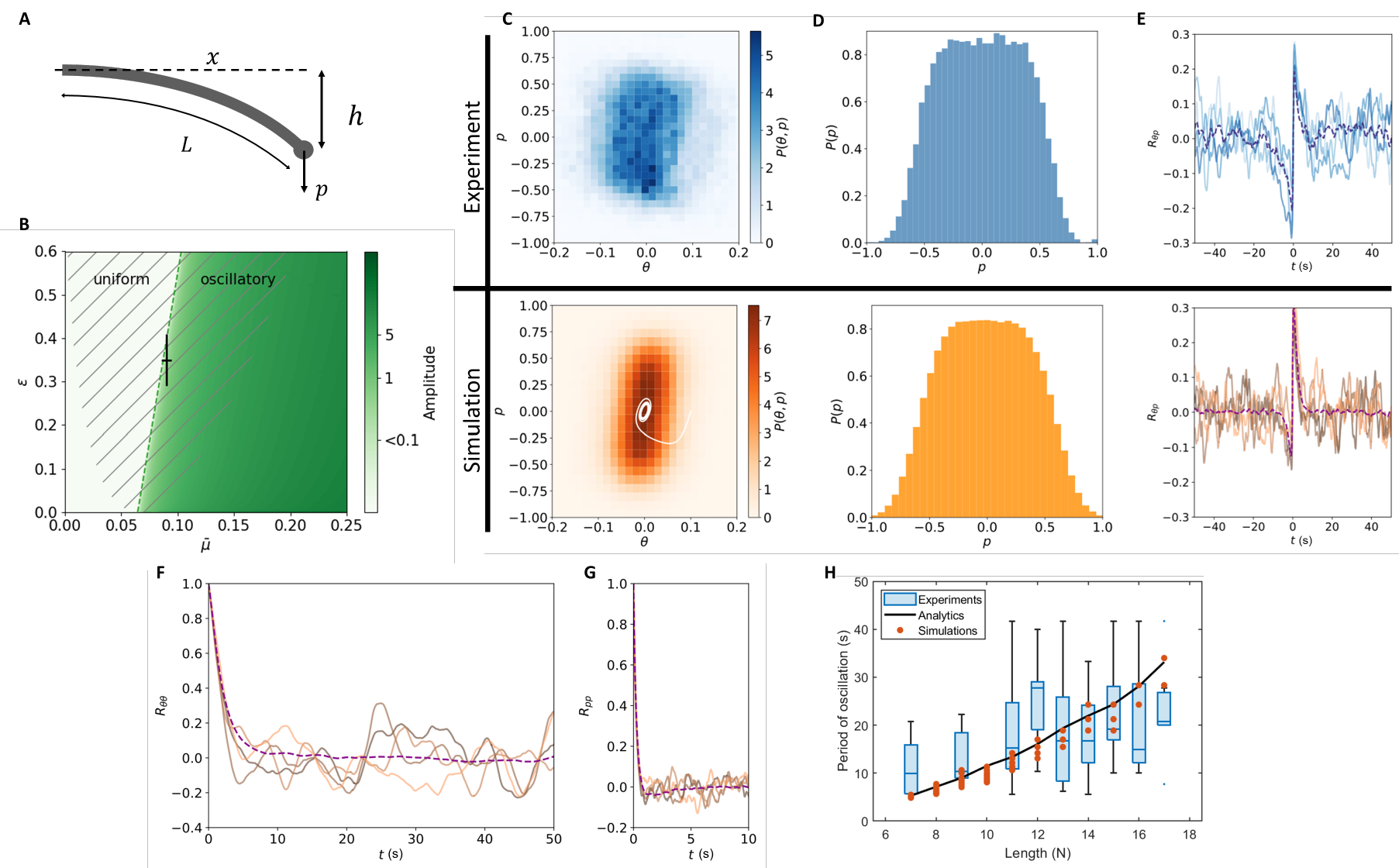}
\caption{ {\bf Theoretical modeling of clamped beams and comparison with the experiment.} This model is obtained by adding the effect of elasticity to the previous model of free beams [see Main text]. (A) Schematics of a clamped beam and definition of the notations. (B) Phase diagram predicted by the model of clamped beams in $\epsilon-\bar{\mu}$ space: the dashed green line marks the onset of oscillations, as predicted by the real part of the largest eigenvalue turning positive, and the shaded region corresponds to  where the same eigenvalue has nonzero imaginary part [SI \JP{sec.~S9}]. The black cross represents the parameters in the experiment from combined estimates [see Main text].  (C-D) Density distribution in ($p, \theta$) space from experiments (blue) and simulations (orange). (E) Cross-correlations $R_{\theta p}$ between  $\theta$ and $p$ in experiments (blue) and simulations (orange). They exhibit excellent agreement between experiments and our model. (F-G) $\theta$ Auto-correlations $R_{\theta\theta}, R_{pp} $ of numerical simulations of parameters predicted by the maximum likelihood estimates of the free beam. (H)  Period of oscillations as a function of length N of active beams and comparison with analytical predictions from the imaginary parts of the largest eigenvalue. In both experiments (blue bars) and theory (black line), only the oscillatory systems are included in the period calculations. For all simulations (red dots) performed, unless stated otherwise, parameters are drawn from Gaussian distributions with mean and standard deviations from row 2 of [Table 2]. }
\label{fig:figure4}
\end{figure*}

\newpage 

\begin{table}[ht!]
    \centering
    \vspace{0.1cm}
    \begin{tabular}{wc{3cm}wc{1.5cm} wc{1.5cm}wc{1.5cm}wc{1.5cm}}
        \toprule
        Method  & $s (s^{-1} ) $ &  $\frac{4}{\gamma} \bar{\alpha}$ & $\frac{4}{\gamma} \beta$ \\ 
        \midrule
        Direct estimates &  $ 17 \pm 2$    &  $699\nu \approx 70$ &  N/A \\
        Stochastic inference  & $55 \pm 3 $ & $81 \pm 16$ & $56 \pm 10$ \\
        \bottomrule
    \end{tabular}
    \caption{{\bf Summary of parameters  for the free beams}. The units are stated in row 1 except for dimensionless parameters. The first row includes parameters estimated from first principles and independent experimental measurements [SI Sec.~S7]. The second row summarizes the best-fit values of the model for free beams using stochastic inference.  The values from the two rows match well with each other.}
    \label{tab:sr}
\end{table}

\begin{table}[htbp]
    \centering
    \vspace{0.1cm}
    \begin{tabular}{wc{3cm}wc{1.8cm} wc{2cm}wc{1.6cm}wc{1.5cm}wc{1.6cm}wc{1.9cm}}
        \toprule
        Method  & $\epsilon (s^{-1}) $ &  $\bar{\mu} (s^{-1}) $ & $\gamma (s^{-1}) $   & $c$ & $\beta (s^{-1}) $  & $\sigma_\theta (s^{-1/2}) $ \\
        \midrule
       First principles & $0.35 \pm  0.04$  & $0.04 \pm 0.01$  & $1 \pm 0.1$ & $\approx 40$ & N/A  &  $0.03 \pm 0.01$  \\
        From free beams & N/A  & $ 0.09 \pm 0.005$  & N/A & $(15.5 \pm 2) \gamma $  & $(15 \pm 2)\gamma $ & N/A \\ 
         \midrule
        Combined estimates & $0.35 \pm  0.04$ & $ 0.09 \pm 0.005$  &  $1 \pm 0.1$ & $15.5 \pm 2$ & $15 \pm 2 $ & $0.03 \pm 0.01$  \\
        \bottomrule
      From clamped beams & $0.1 \pm 0.02$  & $ 0.19 \pm 0.02$  & $0.43 \pm 0.07$ & $5.8 \pm 0.3$ & $12.5 \pm 1 $ & $0.023  $ \\ 
         \bottomrule
    \end{tabular}
    \label{tab:br}
\caption{{\bf Summary of parameters for the clamped beams}. The units are stated in row 1 except for dimensionless parameter $c$. The first row includes parameter estimated from first principles and independent experimental measurements [SI Sec.~S7]. The second row summarizes the best-fit values of the model with the experimental data for free beams [Fig.3G] using stochastic inference. Parameters estimated from first principles or fit of the free beams agree remarkably well.  The third row combines the two rows in order to obtain a full set of parameters to describe the dynamics of the clamped beams.  Those values  match well the best-fit values obtained from the dynamics of clamped beams (Fourth row), thus showing that our model predicts the dynamics of clamped beams from experiments on free beams, which have qualitatively different behavior.} 
     
\end{table}

\newpage
\subsection*{Materials and Methods}
\paragraph{Synthesis of active colloids}
The synthesis of the active colloids is two steps, first synthesizing hematite cubes, then the active colloids. More details are given in \cite{aubret_metamachines_2021}. \\
The synthesis procedure of hematite cubes follows the sol-gel method described by Sugimoto\cite {Sugimoto:1993bf}. Briefly, we mix 100 mL of 2M FeCl$_3$ $\cdot$ 6H$_2$O, 90 ml 6 NaOH and 10 ml water, in a 250 mL pyrex bottle and shake thoroughly. Immediately after, the bottle is placed in an oven at $100^\circ$C, and aged for 3 to 4 days, until the hematite particles reach the desired size (growth is monitored by optical microscopy). The resulting hematite cubes in the gel network are isolated by successive sedimentation and resuspension cycles in DI water [Fig.S2].
The synthesis of the active particles as heterodimers is performed by heterogeneous nucleation of trialkoxysilanes (oil precursors) on solid hematite particles as seeds, and the control of their wetting properties with the pH of the solution. The synthesis procedure is adapted from ref.\cite{Youssef:2016kb}, with chemical modification to reinforce the stability of the active particles under optical trapping. In particular, we make use of a hydrophobic copolymer Hexadecyltrimethoxysilane (HTS) to chemically protect the bond between the hematite and polymer core against highly reactive hydroxil radicals generated during H$_2$O$_2$ consumption. A beaker with 100 mL of DI water is prepared, and mixed with 120 $\mu$L of a 50\% NH$_3$ solution, giving a pH $\sim 10.5$. The solution is kept under mild magnetic stirring. We add $\sim 1$ mL of an aqueous solution of hematite particles, to get a slightly red-colored solution. Following, we add 100 $\mu$L of HTS, immediately followed by 500 $\mu$L of 3-(Trimethoxysilyl)propyl methacrylate (TPM). The solution is then covered with parafilm, and let under mild stirring for $\sim$ 2h00. During this time, the HTS and TPM hydrolysis, and condense on the hematite particles, with strong wetting leading to their engulfment in the oil phase. After $\sim$ 2h00, the solution is turbid. We then add 2 mL of Pluronic F-108 solution (5\% wt), and wait 2 mn. The dewetting and extrusion of TPM from the hematite is performed by changing the pH of the solution to pH $\sim$ 2.1, by adding 1.5 mL of 1M chloridric acid HCl. The solution is let under stirring for 3 mins, and diluted 4 times. We then carry out the polymerization by adding 50 mg of radical initiator Azobisisobutyronitrile (AIBN) to the solution and leave it under stirring for $\sim 5$ mins. The beaker is covered with an aluminium foil and placed in a pre-heated oven at $\sim$ 80 degrees Celsius for 2 hours. We let the solution cool down to room temperature, remove the excess solution above the sedimented particles, and add 50 mL DI water with 1 mL of 250 mM NaOH solution, giving a pH $\sim$ 10, and let the solution overnight to facilitate the hydrolysis and condensation of any remaining HTS monomers. The solution is then centrifuged and rinsed multiple times to remove the excess TPM/HTS particles and obtain the desired colloidal solution of heterodimers. 

\paragraph{Micro-tensile tests}
We trap both ends of an active beam using calibrated optical traps and gradually modulate the distance between the traps until the active beam breaks or exerts forces sufficient to exit the optical traps [SI]. 
We determine the stress acting on the active beam by monitoring the relative displacement $\Delta$x of the active colloids to the center of the traps, while the strain is determined by measuring the total extension $\Delta$L of the active beam with respect to the rest length $L_0$  in the absence of optical traps [SI].  We repeat the  experiment with 8 different beams of the same length N=15  to obtain the stress-strain characterization.
The optical forces of the trap are obtained conventionally by determining the stiffness of the optical trap using the diffusion and the equipartition theorem method [SI].  The positions of the microswimmer positioned in the optical traps are tracked over 4000 frames with a CCD camera.


\section*{Acknowledgments}
The authors thank Andela Saric, Christoph Zechner and Paul Robin for helpful discussions. JP acknowledges  support by ERC grant (VULCAN, 101086998) and US ARO under award W911NF2310008. Views and opinions expressed are however those of the author(s) only and do not necessarily reflect those of the European Union or the European Research Council Executive Agency. Neither the European Union nor the granting authority can be held responsible for them. YIL acknowledges funding from the European Union’s Horizon 2020 research and innovation programme under the Marie Skłodowska-Curie Grant Agreement No. 101034413.

\section*{Authors contributions}
Q.M., A.A. and J.P. conceived the experiments. Q.M. performed the experiment, acquired the data and conducted the data analysis. Y.I.L. and E.H. developed the theoretical model.  Q.M., Y.I.L., E.H and J.P.  interpreted the results and wrote the manuscript. All authors reviewed the manuscript and provided feedback.
\newpage

\bibliographystyle{ieeetr}	
\bibliography{ActiveRodRefJP}

\end{document}